\RequirePackage{fix-cm}

\documentclass[smallextended,authoryear]{svjour3}

\smartqed
\usepackage{dblfloatfix}
\usepackage[section]{placeins}

\usepackage{natbib}
\bibpunct{(}{)}{;}{a}{.}{ }

\usepackage{graphicx}
\usepackage{subcaption}
\usepackage{float}
\usepackage{tikz}
\usepackage{wrapfig}

\usepackage{booktabs}
\usepackage{multirow}
\usepackage{tabularx}
\usepackage{array}

\usepackage[T1]{fontenc}
\usepackage{lmodern}

\usepackage{amsmath}

\usepackage{listings}
\usepackage{minted}

\usepackage{xcolor}
\usepackage[most]{tcolorbox}

\usepackage{hyperref}

\usepackage{xspace}

\newcolumntype{Y}{>{\raggedright\arraybackslash}X}

\definecolor{outergray}{RGB}{244,244,244}
\definecolor{outertitle}{RGB}{170,170,170}

\definecolor{metatitle}{RGB}{225,225,245}
\definecolor{logtitle}{RGB}{245,225,225}
\definecolor{yamltitle}{RGB}{238,238,238}
\definecolor{patchtitle}{RGB}{225,245,225}

\definecolor{boxborder}{RGB}{205,205,205}

\definecolor{yamlgreen}{RGB}{0,120,0}
\definecolor{yamlblue}{RGB}{60,90,160}
\definecolor{yamlred}{RGB}{180,50,50}

\definecolor{diffblue}{RGB}{20,40,160}
\definecolor{diffgreen}{RGB}{0,120,40}
\definecolor{diffred}{RGB}{180,30,30}
\definecolor{diffaddbg}{RGB}{232,247,232}
\definecolor{diffdelbg}{RGB}{252,235,235}

\tcbset{
  exampleouter/.style={
    colback=outergray,
    colframe=boxborder,
    colbacktitle=outertitle,
    coltitle=white,
    boxrule=0.5pt,
    arc=2pt,
    left=6pt,
    right=6pt,
    top=6pt,
    bottom=6pt,
    enhanced,
    width=\textwidth
  },
  metabox/.style={
    colback=white,
    colframe=boxborder,
    colbacktitle=metatitle,
    coltitle=black,
    boxrule=0.4pt,
    arc=2pt,
    left=4pt,
    right=4pt,
    top=4pt,
    bottom=4pt,
    enhanced,
    width=\linewidth
  },
  logbox/.style={
    colback=white,
    colframe=boxborder,
    colbacktitle=logtitle,
    coltitle=black,
    boxrule=0.4pt,
    arc=2pt,
    left=4pt,
    right=4pt,
    top=4pt,
    bottom=4pt,
    enhanced,
    width=\linewidth
  },
  yamlbox/.style={
    colback=white,
    colframe=boxborder,
    colbacktitle=yamltitle,
    coltitle=black,
    boxrule=0.4pt,
    arc=2pt,
    left=4pt,
    right=4pt,
    top=4pt,
    bottom=4pt,
    enhanced,
    width=\linewidth
  },
  patchbox/.style={
    colback=white,
    colframe=boxborder,
    colbacktitle=patchtitle,
    coltitle=black,
    boxrule=0.4pt,
    arc=2pt,
    left=4pt,
    right=4pt,
    top=4pt,
    bottom=4pt,
    enhanced,
    width=\linewidth
  }
}

\lstdefinestyle{instyle}{
  basicstyle=\ttfamily\tiny,
  breaklines=true,
  breakatwhitespace=true,
  columns=fullflexible,
  keepspaces=true,
  showstringspaces=false,
  frame=none,
  aboveskip=2pt,
  belowskip=2pt,
  xleftmargin=0pt,
  xrightmargin=0pt
}

\lstdefinestyle{diffstyle}{
  basicstyle=\ttfamily\tiny,
  breaklines=true,
  breakatwhitespace=true,
  columns=fullflexible,
  keepspaces=true,
  showstringspaces=false,
  frame=none,
  aboveskip=2pt,
  belowskip=2pt,
  xleftmargin=0pt,
  xrightmargin=0pt,
  escapeinside={(*@}{@*)}
}

\newtcolorbox{promptbox}[1][]{
  colback=white,
  colframe=boxborder,
  colbacktitle=outertitle,
  coltitle=white,
  boxrule=0.5pt,
  arc=2pt,
  left=3pt,
  right=3pt,
  top=3pt,
  bottom=3pt,
  enhanced,
  width=\linewidth,
  fonttitle=\bfseries,
  title={Prompt Template},
  #1
}

\newcommand{\benchmark}{{\textsc{CI-Repair-Bench}}\xspace}
\newcommand{\blackcircled}[1]{%
\tikz[baseline=(char.base)]{
  \node[shape=circle,fill=black,inner sep=1.1pt] (char)
  {\textcolor{white}{\footnotesize #1}};
}}

\newcommand{\phead}[1]{\noindent {\bf #1}}

\newcommand{\uheadu}[1]{\noindent\underline{\textit{#1}}}

\begin{document}

\title{CI-Repair-Bench: A Repository-Aware Benchmark for Automated Patch Validation via CI Workflows}

\author{
Rabeya Khatun Muna \and
Md Nakhla Rafi \and
Tse-Hsun (Peter) Chen
}

\institute{
Rabeya Khatun Muna \at
Software Performance, Analysis, and Reliability (SPEAR) Lab, Concordia University, Montreal, QC, Canada \\
\email{rabeyakhatun.muna@mail.concordia.ca}
\and
Md Nakhla Rafi \at
Software Performance, Analysis, and Reliability (SPEAR) Lab, Concordia University, Montreal, QC, Canada \\
\email{mdnakhla.rafi@mail.concordia.ca}
\and
Tse-Hsun (Peter) Chen \at
Software Performance, Analysis, and Reliability (SPEAR) Lab, Concordia University, Montreal, QC, Canada \\
\email{peterc@encs.concordia.ca}
}

\date{Received: date / Accepted: date}

\maketitle

\begin{abstract}
Continuous Integration (CI) enforces repository-level correctness through multi-stage workflows and is central to modern software development, yet diagnosing and repairing CI failures remains challenging. Unlike traditional program repair, CI failures frequently involve non-code artifacts, environment and dependency issues, noisy execution logs, and workflow-level constraints. Existing program repair benchmarks fall short in this setting: they are largely test-centric, restrict repairs to source code, assume fixed execution environments, and evaluate under simplified CI workflows that do not reflect real repository-level validation. We introduce \benchmark, a benchmark for CI-verified, repository-level program repair constructed from real GitHub Actions executions. It contains 567 CI failure instances from 103 repositories and evaluates repair correctness exclusively through full CI re-execution under original workflows. Failures are categorized into 12 CI error types, enabling fine-grained, error-type–aware evaluation. To demonstrate benchmark usage, we include a reference CI repair workflow that analyzes CI logs to localize faults and generate candidate patches. Empirical results show that automated repair is most effective for localized, tool-enforced failures such as formatting and linting, while environment-, dependency-, and configuration-related failures remain challenging; the best-performing LLM achieves an 18.9\% repair success rate. \benchmark provides a realistic evaluation foundation for advancing research on CI-native automated program repair.
\end{abstract}


\section{Introduction}
\label{sec:introduction}
Continuous Integration (CI) is a core mechanism in modern software engineering that aims to enforce repository-level correctness under frequent code and dependency changes. In CI-oriented workflows, each code change triggers a multi-stage validation pipeline executed across isolated environments prior to integration. Unlike standalone testing, CI pipelines simultaneously validate environment setup, dependency resolution, build configuration, static analysis, and test execution, making CI outcomes inherently sensitive to both code and non-code artifacts~\citep{saavedra2024gitbugactions,zhang2024actionsremaker}.

Despite the high degree of automation, CI pipelines frequently fail in practice.
Such failures arise from a diverse range of causes, including configuration
errors, dependency issues, environment drift, coding style violations, and test
failures, often blocking development progress until they are resolved
\citep{zhang2024actionsremaker,moriconi2025ghalogs}. Therefore, diagnosing CI failures is a recurring and costly maintenance task that involves more than inspecting execution logs in isolation: developers must use verbose and noisy CI
logs as an entry point and reason about the broader context, including the
execution environment, recent code changes, library dependencies, and workflow
structure, to accurately identify failure causes
\citep{brandt2020logchunks,xu2024logsage}. 

Unlike traditional program repair,
which typically focuses on localized source code changes under fixed execution
environments, CI issue repair requires reasoning across heterogeneous artifacts
and execution stages, as failures may occur at any point in the pipeline and often involve artifacts beyond source code, such as build scripts, dependency
specifications, environment variables, and workflow configurations
\citep{saavedra2024gitbugactions,zhang2024gha-talk}. Consequently, effective CI issue repair requires understanding CI pipeline execution behavior and its interactions with the surrounding environment and project structure that are not captured by existing benchmarks for automated code generation and program repair.

Most benchmarks for automated program repair rely on simplified verification settings. While effective for controlled evaluation, such settings overlook CI repair scenarios that often involve non-code artifacts, environment-dependent failures, cross-file interactions, and multi-stage verification at the repository level. Existing benchmarks typically exhibit one or more of the following limitations:

\blackcircled{1}\ \textbf{Repair scope is limited to source code artifacts.}
Most program repair and code generation benchmarks restrict candidate repair actions to program source files. Existing benchmarks~\citep{sun2024repobench,zhang2023repocoder,jimenez2024swebench,chen2021evaluating,austin2021mbpp} typically do not permit modifications to non-code artifacts, including build scripts, dependency specifications, environment variables, or CI workflow configurations. Consequently, CI failures that require changes beyond source code cannot be addressed within these benchmark settings.

\blackcircled{2}\ \textbf{Quality is evaluated using test-centric oracles.}
Prior benchmarks typically evaluate repair correctness using test execution as the sole oracle, treating test pass or fail as the only indicator of success~\citep{chen2021evaluating,austin2021mbpp,sun2024repobench,yang2024execrepobench,jimenez2024swebench}. This test-centric evaluation overlooks additional constraints enforced in CI pipelines, including formatting rules, static analysis, configuration validity, and workflow execution behavior~\citep{saavedra2024gitbugactions,zhang2024actionsremaker}.

\blackcircled{3}\ \textbf{Execution environment and dependency failures are not considered.}
Most program repair benchmarks, such as the SWE-Bench family~\citep{jimenez2024swebench,xie2025swebenchplus,jimenez2024swebenchpro,openai2024swebenchverified}, evaluate patches under fixed or containerized execution environments, implicitly assuming stable dependencies, toolchains, and system libraries. As a result, failures arising from dependency resolution, environment provisioning, or tooling drift are excluded from evaluation.

\blackcircled{4}\ \textbf{Failure diagnosis is issue-centric rather than log-driven.}
Issue-driven benchmarks, such as the SWE-bench family, rely on bug reports or issue descriptions and assume that fault localization information is available. In contrast, CI failures are commonly diagnosed using long and noisy execution logs together with recent code change context, leaving a gap between CI failure diagnosis and executable repair evaluation~\citep{brandt2020logchunks,xu2024logsage,zhang2024gha-talk}.

\blackcircled{5}\ \textbf{Verification is limited to a subset of CI workflow stages at scale.}
Existing CI-aware benchmarks evaluate repairs on a limited number of repositories and simplified workflow configurations~\citep{saavedra2024gitbugactions,bogomolov2024longcodearena}. As a result, they fail to capture the diversity of real-world CI pipelines, including conditional execution, repeated validation across environments, and project-specific validation stages~\citep{zhang2024actionsremaker}.

\begin{table}[t]
\caption{Comparison of repair benchmarks by repair target, validation oracle, and CI verification scope.}
\label{tab:ci_benchmark_comparison}
\centering
\scriptsize
\renewcommand{\arraystretch}{1.15}
\setlength{\tabcolsep}{4pt}

\resizebox{\columnwidth}{!}{
\begin{tabular}{p{1.5cm} p{2.1cm} p{0.8cm} p{2.8cm} p{1.8cm} p{2.8cm}}
\toprule
\textbf{Benchmark} &
\textbf{Repair Target} &
\textbf{Size} &
\textbf{Validation Oracle} &
\textbf{Execution Semantics} &
\textbf{CI Verification Scope} \\
\midrule

\multicolumn{6}{l}{\textbf{General Issue Repair (Test-Centric)}} \\

SWE-bench
& Issue-level code repair
& 2,294
& Test pass/fail
& Containerized
& Unit and integration tests \\

SWE-bench+
& Issue-level code repair
& 548
& Test pass/fail (+ manual checks)
& Containerized
& Unit and integration tests \\

SWE-bench Verified
& Issue-level code repair
& 500
& Test pass/fail
& Containerized
& Unit and integration tests \\

SWE-PolyBench
& Issue-level code repair
& 2,110
& Test pass/fail
& Containerized
& Unit and integration tests \\

SWE-bench Pro
& Issue repair and feature addition
& 1,865
& Human-verified outcome
& Containerized
& Unit and integration tests \\

FEA-Bench
& Feature addition
& 1,401
& Test pass/fail
& Containerized
& Unit and integration tests \\

\midrule
\multicolumn{6}{l}{\textbf{CI Repair}} \\

GitBug-Actions
& Workflow-level CI repair
& 21
& (1) Test execution
& GitHub Actions (Act)
& (1) Single CI job \newline
  (2) Build and test stages \\

LCA-CI-Repair
& Workflow-level CI repair
& 77
& (1) Static analysis \newline
  (2) Test execution
& GitHub Actions (hosted)
& (1) Single workflow \newline
  (2) Multiple CI jobs \newline
  (3) Repeated execution across environments \\

\textbf{CI-REPAIR-BENCH}
& \textbf{Repository-level CI repair}
& \textbf{567}
& \textbf{(1) Build execution \newline
          (2) Dependency resolution \newline
          (3) Configuration validity \newline
          (4) Static analysis \newline
          (5) Test execution}
& \textbf{GitHub Actions (hosted)}
& \textbf{(1) Multiple interdependent CI jobs \newline
          (2) Conditional execution paths \newline
          (3) Repeated execution across environments or configurations \newline
          (4) Nested or reusable workflow execution} \\

\bottomrule
\end{tabular}
}
\end{table}

To address these limitations, we introduce \textbf{CI-REPAIR-BENCH}, a benchmark of real CI failures collected from GitHub repositories that is designed to reflect realistic CI repair settings. CI-REPAIR-BENCH supports \blackcircled{1} \emph{repository-level repair}, allowing fixes to both source code and non-code artifacts such as dependency specifications, configuration files, and CI workflow definitions. \blackcircled{2} Repair correctness is evaluated via \emph{full GitHub Actions re-execution} rather than test-centric oracles, enforcing all CI validation stages, including formatting, linting, configuration checks, environment setup, and test execution under native CI settings. \blackcircled{3} The benchmark preserves complete CI pipelines, including dependency resolution, configuration logic, environment provisioning, and multi-stage validation, enabling analysis of failures beyond source code rather than filtering them out during dataset construction. \blackcircled{4} Each instance is log and code-history driven rather than issue-driven, providing long and noisy CI execution logs as the primary diagnostic signal and reflecting real-world CI debugging practice. Finally, \blackcircled{5} CI-REPAIR-BENCH includes complex CI workflows with multiple jobs, going beyond simplified or single-step pipelines.

To illustrate how CI-REPAIR-BENCH can be used to evaluate automated repair under realistic CI validation, we introduce a reference CI repair framework that integrates CI log analysis and fault localization to generate repository-level patches from failed CI executions. Evaluated on all 567 benchmark instances under full CI re-execution, the framework achieves repair success rates of up to 18.9\%. Repair effectiveness varies substantially across CI failure categories and model capabilities: formatting and static analysis failures are often resolved, whereas failures involving dependencies, environment setup, and workflow logic remain largely unsolved. These results highlight both the potential and current limitations of automated CI repair under realistic validation and motivate further research on CI-aware repair methods.

Overall, our contributions in this work are threefold:

\begin{itemize}

\item We introduce \benchmark, a large-scale benchmark for
\textbf{CI-validated, repository-level program repair} constructed from real GitHub Actions executions. Each instance captures a complete CI failure episode, including workflow configurations, execution logs, repository metadata, fail-to-pass commit pairs, and developer-written fixes. Repair correctness is evaluated via \textbf{full GitHub Actions re-execution}, enabling realistic assessment under native CI semantics.

\item We present a \textbf{reference CI repair pipeline} that integrates \emph{CI log
analysis}, \emph{fault localization}, and \emph{patch generation} to demonstrate how automated repair methods can be evaluated on \benchmark. The pipeline leverages CI logs and failing-commit context to localize faults and generate repository-level patches using LLM-based models, whose correctness is assessed through complete CI re-execution. We instantiate this pipeline with four LLMs to study how model choice impacts CI repair performance.

\item We conduct an \textbf{error-type--aware empirical study of LLM-based CI repair} using
\benchmark. Our analysis characterizes how repair effectiveness varies across distinct CI failure modes, showing that current LLM-based methods are effective for a subset of tool-enforced failures, such as formatting and linting, but struggle with failures involving dependencies, environment setup, configuration, and workflow logic.

\end{itemize}

\phead{Paper Organization.} Section~\ref{sec:relatedwork} discusses background and related work. Section~\ref{sec:benchmark} presents CI-Repair-Bench, including task definition, benchmark construction, benchmark maintenance, and dataset summary. Section~\ref{sec:experiment} presents the reference CI repair framework. Section~\ref{sec:result} reports the experimental results. Section~\ref{sec:discussion} discusses the implications, limitations, and challenges of CI repair. Section~\ref{sec:threats} discusses threats to validity. Finally, Section~\ref{sec:conclusion} concludes the paper.

\section{Related Work}
\label{sec:relatedwork}

Continuous Integration failures pose recurring challenges in modern software development, while automated program repair has emerged as a promising approach for reducing the effort required to diagnose and fix software defects. We organize the related work as follows: first, studies on CI-centric benchmarks and datasets that leverage CI workflows, execution logs, and pipeline outcomes for empirical analysis or repair evaluation; second, automated program repair (APR) benchmarks that evaluate patch generation for real software defects under executable settings.

\subsection{{CI-centric Benchmarks and Datasets}} 
Table~\ref{tab:ci_benchmark_comparison} summarizes benchmarks that leverage continuous integration systems to study real-world software engineering. Early CI-centric datasets such as GHALogs~\citep{moriconi2025ghalogs} and BugSwarm~\citep{tomassi2019bugswarm} collect CI workflows and execution logs to support empirical analyses of CI usage and failure characteristics. However, these datasets are primarily descriptive. They do not define executable repair tasks for end-to-end evaluation of automated CI repair.

More recent benchmarks incorporate CI execution as a correctness signal. GitBug-Actions replays GitHub Actions workflows and treats CI success as an oracle~\citep{saavedra2024gitbugactions}, while ActionsRemaker reconstructs execution environments to enable offline workflow reproduction~\citep{zhang2024actionsremaker}. Long Code Arena includes CI build repair as a subtask within a broader evaluation suite~\citep{bogomolov2024longcodearena}. However, these benchmarks remain limited in scale or scope, often evaluate simplified or isolated workflows, and lack repository-wide CI coverage or explicit failure-type annotation.

In contrast, \benchmark\ evaluates automated repair by re-executing complete GitHub Actions workflows under native CI semantics, defining correctness at the pipeline level across all jobs and dependencies, and enabling error-type-aware analysis of repository-level CI repair.

\subsection{{Automated Program Repair (APR) Benchmarks}}
Automated Program Repair (APR) benchmarks evaluate systems that generate patches for real software defects. Early benchmarks such as Defects4J focus on unit-level bug fixes revealed by test failures~\citep{just2014defects4j,defects4j_cleanness_2024}. 
More recent benchmarks move toward repository-level, issue-driven repair. The SWE-bench family evaluates issue resolution as a standard task, with later variants improving data quality and reducing test leakage~\citep{jimenez2024swebench,xie2025swebenchplus}. SWE-bench Pro further increases task complexity by introducing long-horizon, multi-file repairs drawn from large-scale repositories~\citep{jimenez2024swebenchpro}. These benchmarks expand APR evaluation from unit-level bug fixes to repository-level, issue-driven repair with longer execution horizons.

Nevertheless, existing APR benchmarks remain fundamentally test-centric and are evaluated under fixed, containerized execution environments. This design limits their ability to capture CI-specific failure modes that arise outside test execution, such as workflow misconfiguration, dependency resolution failures, environment drift, or pipeline orchestration errors. In contrast, our benchmark evaluates repository-level repair by re-executing complete CI workflows on GitHub Actions. This enables pipeline-level validation that reflects CI behavior beyond test outcomes alone.

\section{\benchmark}
\label{sec:benchmark}

\begin{figure}[t]
  \centering
  \includegraphics[width=0.9\columnwidth]{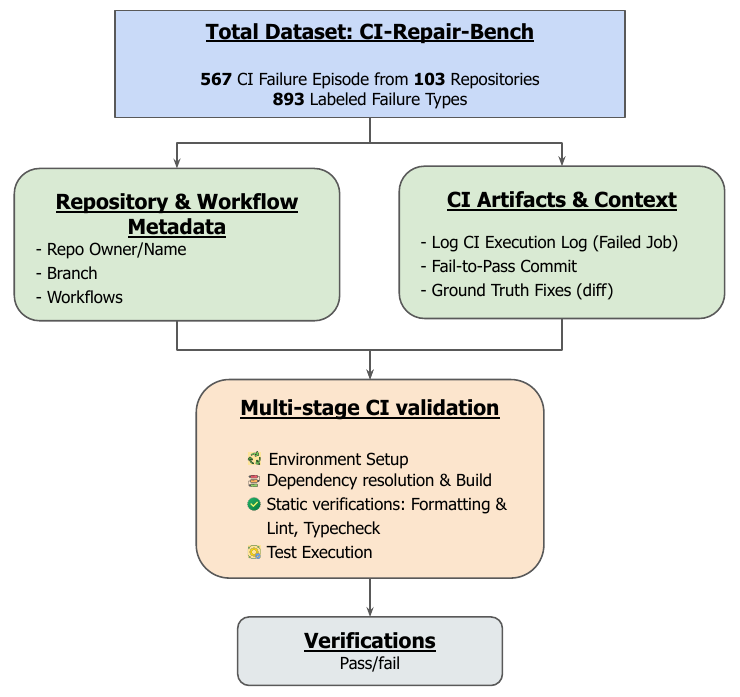}
  \caption{Overview of \benchmark.}
  \label{fig:data-overview}
\end{figure}

Figure~\ref{fig:data-overview} presents an overview of \benchmark, illustrating both the structural components of a benchmark instance and the CI validation context used for evaluation. Each instance consists of five components:
\blackcircled{1} \textbf{Repository Metadata}: specifies the repository owner, repository name, and branch required to reconstruct the project state;
\blackcircled{2} \textbf{CI Workflow Details}: describes the CI workflow associated with the failure, including its identifier, configuration file path, and the validation steps executed during CI;
\blackcircled{3} \textbf{CI Logs and Error Artifacts}: execution logs, error messages, and test or analysis outputs produced by the CI workflow, which provide detailed signals for diagnosing failures and evaluating candidate repairs.
\blackcircled{4} \textbf{Failure-to-Success Commit Pair}, consisting of a failing commit and its nearest subsequent passing commit on the same branch under the same CI workflow;
\blackcircled{5} \textbf{Ground-Truth Patch}: a repair-relevant patch with the minimal set of changes necessary to resolve the CI failure, derived from the commit pair by retaining only causally related modifications that are validated through CI re-execution.

\subsection{Task Definition and Evaluation Metrics}
Figure~\ref{fig:data-overview} illustrates the task and evaluation workflow of \benchmark. 
Given a failing commit, the corresponding repository state, and the associated CI workflow configuration, the task is to generate a candidate patch that repairs the CI failure. The evaluation is conducted through a standardized execution pipeline that applies the generated patch and re-executes the original CI workflow under identical conditions using GitHub Actions. A repair is considered successful if and only if the CI workflow transitions from failure to success without introducing new CI failures.

\phead{Evaluation Setup and Execution.}
Given a generated patch, the benchmark follows the automated evaluation steps below.

\uheadu{(1) Repository Setup.}
For each benchmark instance, the repository is cloned and checked out to the corresponding failing commit. A dedicated benchmark branch is then created to isolate evaluation and ensure reproducibility across instances.

\uheadu{(2) Workflow Selection and Standardization.}
The CI workflow associated with each failure is selected and standardized to enable efficient and consistent re-execution while preserving validation semantics. For each instance, we retain only the failing workflow and discard unrelated workflows to avoid unnecessary execution overhead. The original workflow is transformed into a functionally equivalent, simplified version triggered by a single \texttt{push} event, where redundant execution dimensions (e.g., duplicated operating system or Python version matrices) and non-validation steps (e.g., artifact publishing or release-related actions) are removed. All steps responsible for correctness verification, including environment setup, dependency installation, build, static analysis, and test execution, are preserved without modification to commands, tool configurations, environment variables, or job dependencies. To ensure semantic equivalence, we re-execute the developer-authored fix under the standardized workflow and retain only instances for which the CI outcome (pass or fail) matches that of the original workflow; instances that violate this equivalence criterion are excluded. These safeguards ensure that workflow standardization improves execution efficiency and reproducibility without altering the original verification behavior.

\uheadu{(3) Patch Application and CI Re-execution.}
Each generated patch is first checked for applicability by applying it to the repository at the failing commit. A patch is considered applicable if it applies cleanly without conflicts or syntax errors and modifies only files that exist in the repository at that commit. Only applicable patches are used to re-execute the CI workflow under the original conditions via GitHub Actions.

\uheadu{(4) CI Outcome Collection.} CI workflow execution is monitored until completion, and the benchmark records the final execution status of each job to determine a workflow-level outcome. A repair is considered \textit{successful} if and only if all jobs complete successfully for the applied patch; otherwise, it is considered \textit{failed}. Beyond this binary outcome, successful workflow executions are mapped back to the CI error-type labels associated with each instance, enabling analysis of which failure categories are resolved overall by a given repair.

\phead{Evaluation Metrics.}
\benchmark evaluates candidate patches using Pass@1 (i.e., each benchmark instance is evaluated using a single candidate patch) using two metrics: \textit{repair success rate} and \textit{per-error-type repair accuracy}. 

\uheadu{(1) Repair Success Rate.}
Let $\mathcal{D}$ denote the full benchmark dataset, and let $\mathcal{S} \subseteq \mathcal{D}$ denote the subset of instances whose CI workflows pass after patch application. The repair success rate is defined as:
\[
\text{SuccessRate} =
\frac{|\mathcal{S}|}{|\mathcal{D}|} \times 100.
\]
Instances for which CI re-execution cannot be successfully triggered are counted as unsuccessful repairs. This metric measures end-to-end CI repair effectiveness under repository-level execution.

\uheadu{(2) Per-Error-Type Repair Accuracy.}
For each CI error category, we report the repair success rate computed over instances annotated with that category, enabling analysis of repair effectiveness across different types of CI failures.

\subsection{Automated Benchmark Construction Pipeline}
\label{sec:dataset_construction_pipeline}

We construct \benchmark using an automated pipeline that collects real-world CI failure instances from GitHub Actions while enforcing reproducibility, re-executability, and consistent evaluation conditions. The dataset construction proceeds in five steps.

\textbf{Step 1: Repository Selection.}
We select open-source Python repositories that use GitHub Actions. We focus on Python because it has a large and actively maintained open-source ecosystem with mature CI workflows on GitHub Actions spanning testing, linting, formatting, dependency management, and environment validation. Repository selection follows two criteria:
(i) at least 1{,}000 stars and forks, indicating mature development practices~\citep{kalliamvakou2014promises}; and
(ii) development activity within the past 90 days at the time of data collection (early 2025).
This activity constraint ensures that CI pipelines are actively exercised and that execution logs remain accessible, as GitHub Actions logs are removed after 90 days by default~\citep{bogomolov2024longcodearena}.
After applying these criteria and subsequent filtering steps, 103 repositories are retained in the final benchmark.
\textbf{Step 2: CI Run Collection and Pair Extraction.}
For each repository, we retrieve the GitHub Actions execution history and extract failure--success commit pairs following established practices for mining bugs and fixes~\citep{tomassi2019bugswarm,bogomolov2024longcodearena}.
A pair consists of a failing commit and the nearest subsequent commit that triggers a successful run of the same workflow on the same branch.
We retain only pairs where
(i) both runs correspond to the same workflow,
(ii) at least one modified file is a Python source file, and
(iii) workflows unrelated to program validation (e.g., deployment, release automation, or infrastructure maintenance) are excluded.

\textbf{Step 3: Data Cleaning and Filtering.}
We apply a series of filters to ensure reliable CI re-execution and fair evaluation. First, automated filtering removes workflows whose names commonly indicate non-program-level CI intent (e.g., release, deployment, docker, dependabot, or documentation updates).
We further exclude failure--success pairs with large changes (more than 15 modified files) without containing any Python source files, as these instances are unlikely to reflect program-level CI validation failures in Python projects and often arise from unrelated large-scale edits.

The remaining workflows are further manually examined to remove cases for which faithful re-execution is not feasible in public or forked environments, such as those requiring self-hosted runners, organization-restricted infrastructure, deprecated runners, unsupported third-party services, or access to confidential secrets. In addition, instances showing unstable CI outcomes due to dependency drift, service unavailability, or non-deterministic workflow behavior are excluded to preserve benchmark reproducibility.

Finally, workflow configurations are standardized by removing non-validation steps and reducing redundant execution matrices while preserving verification semantics. All validation-related steps, including environment setup, dependency installation, static analysis, and test execution, are preserved to ensure consistent CI behavior across instances.

\textbf{Step 4: Ground-Truth Patch Extraction.} The developer-authored passing commit is not directly used as the ground-truth repair, since it may contain unrelated edits such as refactoring, formatting changes, feature additions, dependency updates, or repository-wide maintenance modifications. Instead, we perform failure-aware change extraction over the fail-to-pass commit pair to isolate only the modifications directly relevant to the observed CI failure. This process is guided by the failed job logs, workflow validation steps, and changed artifacts. The extracted repair-relevant changes are incrementally applied to the failing commit and validated through full CI re-execution. Only the minimal validated set of modifications that restores the original CI workflow is retained as the ground-truth patch.

\textbf{Step 5: Dataset Assembly and Validation.} For each retained instance, we assemble a complete benchmark record comprising repository metadata (repository name, owner, and branch), the failing and passing commit identifiers, the CI workflow configuration at the failing commit, the corresponding failed job logs, and the validated ground-truth patch. Each instance is further annotated with one or more CI error-type labels derived from the CI failure context and the validated repair. As CI workflows often involve multiple validation stages, a single instance may be associated with multiple CI error categories. To ensure benchmark reliability, we validate each retained instance by re-executing both the failing commit and the validated ground-truth patch under the standardized workflow. Instances exhibiting inconsistent CI outcomes, flaky executions, or newly introduced unrelated failures are excluded, ensuring that all retained benchmark instances correspond to stable and reproducible CI failure episodes.

\subsection{Benchmark Maintenance}
\label{sec:benchmark_maintenance}

CI workflows evolve over time due to deprecated package versions, runner updates, unavailable services, and changes in external dependencies, which may render previously valid benchmark instances irreproducible. To ensure long-term reproducibility, all retained instances undergo periodic re-validation under semantics equivalent to the original workflow intent. When workflow components become deprecated (e.g., unavailable package versions, deprecated runners, obsolete workflow actions, or unsupported external services), we update the workflow to the nearest reproducible configuration while maintaining the original validation stages and toolchain behavior. If such updates introduce additional validation failures, we re-execute the workflow and incorporate only the minimal additional code changes required to restore end-to-end CI success into the ground-truth patch. In these cases, the workflow configuration, failed job logs, validation artifacts, and ground-truth patch are updated accordingly while preserving the original failure category and repair intent.

\subsection{Dataset Summary}
\label{ref:dataset_summary}

The final benchmark consists of 567 CI failure instances collected from 103 actively maintained Python repositories. Each instance is constructed from a failure--success commit pair executed under the same CI workflow: the failing commit captures the original CI failure, while the succeeding commit corresponds to the developer-authored repair. This construction preserves the original repository state and execution context, enabling realistic and reproducible evaluation. Each instance includes repository metadata, CI workflow configuration, execution logs, the failure--success commit pair, and a validated ground-truth patch that isolates the repair-relevant changes.

In total, the benchmark contains 893 CI error-type labels across 567 instances, reflecting the multi-stage and potentially multi-failure nature of real-world CI workflows. These labels span 12 failure categories, covering both localized issues (e.g., linting, formatting, and testing errors) and system-level failures (e.g., dependency, configuration, and environment issues). All instances are validated through repeated CI re-execution under standardized workflows while preserving the original verification semantics, ensuring the reliability of both failure reproduction and repair validation.

To illustrate the structure of a benchmark instance, Figure~\ref{fig:example_instance} presents a representative example. The figure highlights how repository metadata, workflow configuration, CI failure signals, and the corresponding ground-truth patch are jointly captured within a single instance, providing a complete view of the failure and its resolution.

\begin{figure*}[t]
\centering
\footnotesize

\begin{tcolorbox}[exampleouter,title={Example \benchmark Instance (Formatting Failure)}]

\begin{minipage}[t]{0.48\textwidth}
\begin{tcolorbox}[metabox,title={Metadata}]
\begin{tabularx}{\linewidth}{@{}lY@{}}
\textbf{language:} & Python \\
\textbf{id:} & 1 \\
\textbf{repo\_owner:} & huggingface \\
\textbf{repo\_name:} & diffusers \\
\textbf{head\_branch:} & ipadapterfaceid \\
\end{tabularx}
\end{tcolorbox}
\end{minipage}
\hfill
\begin{minipage}[t]{0.48\textwidth}
\begin{tcolorbox}[yamlbox,title={Workflow (YAML excerpt)}]
\begin{lstlisting}[style=instyle]
runs-on: ubuntu-latest
steps:
  - uses: actions/checkout@v3
  - uses: actions/setup-python@v4
  - run: pip install .[quality]
  - run: ruff format --check
\end{lstlisting}
\end{tcolorbox}
\end{minipage}

\vspace{4pt}

\begin{tcolorbox}[metabox,title={Workflow and Instance Details}]
\begin{tabularx}{\linewidth}{@{}lY@{}}
\textbf{workflow\_name:} & Run code quality checks \\
\textbf{workflow\_filename:} & pr\_quality.yml \\
\textbf{workflow\_path:} & .github/workflows/pr\_quality.yml \\
\textbf{sha\_fail:} & 2c06ffa4...f214ff9 \\
\textbf{sha\_success:} & 217d9d07...8c1449 \\
\textbf{changed\_files:} & examples/.../ip\_adapter\_face\_id.py \\
\textbf{error\_type:} & [Code Formatting] \\
\end{tabularx}
\end{tcolorbox}

\vspace{4pt}

\begin{minipage}[t]{0.48\textwidth}
\begin{tcolorbox}[logbox,title={Logs (\{step, log\})}]
\begin{lstlisting}[style=instyle]
{
  "step": "check_code_quality",
  "log": "examples/...:15:1: I001.Import block is un-sorted or un-formatted..Found 1 error."
}
\end{lstlisting}
\end{tcolorbox}
\end{minipage}
\hfill
\begin{minipage}[t]{0.48\textwidth}
\begin{tcolorbox}[patchbox,title={Ground-Truth Patch}]
\begin{lstlisting}[style=diffstyle]
diff --git a/examples/community/ip_adapter_face_id.py b/examples/community/ip_adapter_face_id.py
index d9325742c..e3c5a2c84 100644
@@ -14,12 +14,12 @@
(*@\colorbox{diffaddbg}{\textcolor{diffgreen}{+ from safetensors import safe\_open}}@*)
 ...
(*@\colorbox{diffdelbg}{\textcolor{diffred}{- from safetensors import safe\_open}}@*)
\end{lstlisting}
\end{tcolorbox}
\end{minipage}

\end{tcolorbox}

\caption{Example \benchmark instance illustrating repository metadata, workflow configuration, CI failure signal, and the validated ground-truth patch.}
\label{fig:example_instance}
\end{figure*}

At the dataset level, Table~\ref{tab:repo_stats} summarizes the per-repository distribution of instances and associated error types. The distribution is highly skewed: a small number of repositories account for a large proportion of failures, while many repositories contribute only a few instances. This skew reflects the nature of CI data collection, where repositories with more frequent or stricter validation pipelines naturally produce more repairable failures.

Finally, Table~\ref{tab:ci_failure_by_model_two_strategies} provides an overview of failure-type distribution across the benchmark, capturing the diversity of CI failure categories and their relative prevalence.

\section{Reference CI Repair Framework}
\label{sec:experiment}

\begin{figure*}[t]
  \centering
  \includegraphics[width=\textwidth]{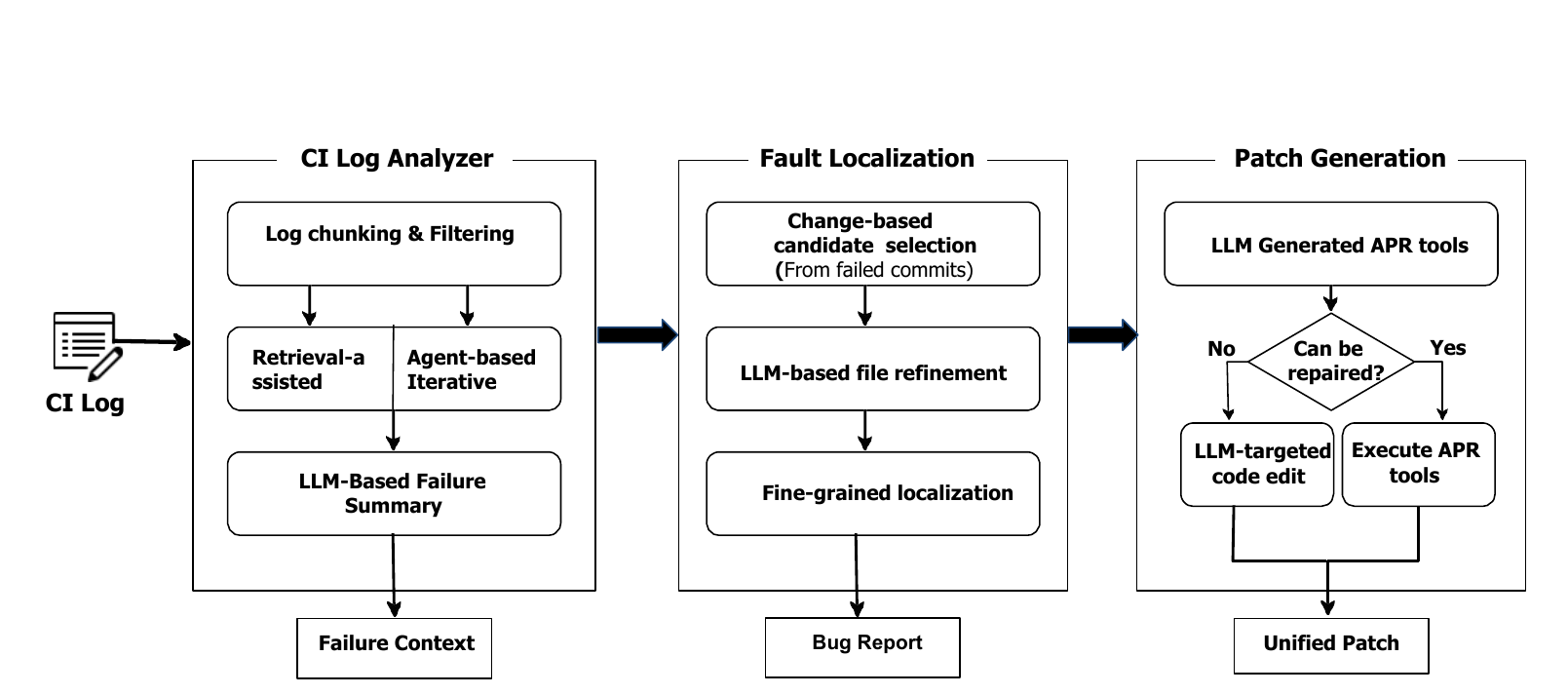}
  \caption{Overview of the LLM-based CI repair pipeline used for evaluation in \benchmark, comprising CI log analysis, fault localization, and patch generation, with repair correctness validated via full CI re-execution.}
  \label{fig:baseline-overview}
\end{figure*}
We define a reference CI repair framework that operationalizes \benchmark for evaluating automated CI repair under realistic validation conditions. 
Given a failing benchmark instance consisting of the repository state at the failing commit, the associated CI configuration, and execution logs from the failed run, the framework applies a fixed three-stage pipeline comprising \emph{(1) CI log analysis}, \emph{(2) fault localization}, and \emph{(3) patch generation}. 
The framework produces a single candidate patch, which is validated via full CI re-execution, as illustrated in Figure~\ref{fig:baseline-overview}. 

\subsection{CI Log Analysis}
CI logs are often long and noisy, spanning multiple execution stages and tools, which can exceed LLM context limits and obscure failure-relevant signals. The goal of this stage is to extract a concise and structured failure summary for downstream localization and repair.

\phead{Token-based log chunking and filtering.}
CI logs are partitioned into token-bounded segments while preserving execution order. Each segment is constrained relative to the model context window to ensure sufficient capacity for reasoning and output generation. When the number of segments exceeds a fixed budget, only a subset is selected: the most recent segments are prioritized, as they typically contain failure-relevant signals, while earlier segments are included only if they contain explicit failure indicators such as error messages, exceptions, or non-zero exit codes. This filtering reduces noise while preserving critical diagnostic information.

\phead{Log analysis strategies.}
We use an agent-based iterative strategy for analyzing filtered log segments within the reference framework.

\uheadu{Agent-based iterative log analysis.}
The LLM analyzes log segments iteratively while maintaining an intermediate structured state, incrementally refining a failure summary by focusing on evidence that explains the observed errors. Intermediate summaries are accumulated and re-analyzed to construct a global failure representation. The iteration budget and output format are fixed across all experiments.

\phead{Prompt design.}
The LLM is guided by a constrained prompt to produce structured failure summaries grounded strictly in log evidence.

\begin{promptbox}[title={Prompt Template: Iterative Log Analysis (Simplified)}]
\begin{verbatim}
Task: Analyze CI logs from a failed run.

Instructions:
1. Extract key failure signals and their surrounding lines.
2. Extract file paths mentioned in logs.
3. Avoid speculation beyond log evidence.

Output:
JSON with summary and relevant files.
\end{verbatim}
\end{promptbox}

\subsection{Fault Localization}
Using the structured failure summary produced by the log analysis stage, the framework identifies files and code regions most likely responsible for the CI failure. Rather than analyzing the entire repository, the fault localization stage progressively narrows the search space in three steps.

(1) \emph{Change-based candidate selection}: files modified in the failing commit and its recent history are collected as initial candidates, based on prior evidence that CI failures are often introduced by recent changes~\citep{chen2022useful}. Commit traversal continues until a successful CI run is encountered or CI metadata becomes unavailable.

(2) \emph{LLM-based file refinement}: each candidate file is evaluated using the failure summary, CI context, and the file’s recent changes. Files are retained only when there is explicit causal evidence linking them to the observed failure, reducing false positives introduced by superficial correlations.

(3) \emph{Fine-grained localization}: for each retained file, a structural outline (e.g., functions, classes, imports) is constructed, and the file is analyzed in bounded segments guided by the failure summary. For each segment, the LLM produces a structured localization report specifying affected line ranges, issue type, and justification grounded in CI evidence.

\begin{promptbox}[title={Prompt Template: Suspicious File Selection}]
\begin{verbatim}
Task:
Determine whether a changed file is responsible for a CI 
failure.

Inputs:
- File path
- Unified diff for the failed commit
- Description of failed CI jobs

Instructions:
1. Decide if the file changes are directly related to the CI 
   failure.
2. Mark the file as suspicious only if there is strong 
   evidence from the diff and CI context.
3. Do not speculate or infer indirect relevance.

Output:
{ "is_suspicious": true | false }
\end{verbatim}
\end{promptbox}

\begin{promptbox}[title={Prompt Template: Fault Localization on Candidate Files}]
\begin{verbatim}
Task:
Identify all failure-relevant faults within a source code file.

Inputs:
- Numbered source code chunks
- File outline (functions, classes, imports)
- CI error context and failed jobs

Instructions:
1. Scan all lines in the given chunk exhaustively.
2. Report only faults directly supported by CI evidence.
3. Precisely localize faults using exact line ranges.
4. Classify each fault by scope (line, method, class, 
   import block).
5. Do not speculate or expand beyond the provided code.

Output:
JSON array of fault objects:
- line_range
- reason
- issue_type
- fault_localization_level
\end{verbatim}
\end{promptbox}

\subsection{Patch Generation}
The final stage produces a repository-level patch intended to repair the CI failure at the failing commit, using localized fault information and CI workflow context.

Patch generation proceeds in three steps.

(1) \emph{Tool-aware repair}: if the localized failure corresponds to a rule violation enforced by a tool already used in the CI workflow (e.g., formatting or linting), the LLM generates installation and execution commands for that tool, enabling deterministic fixes.

\phead{Repair strategy selection.}
Deterministic tool-based repair is prioritized when applicable. If no suitable tool applies, the framework falls back to LLM-based code generation constrained to localized fault regions.

(2) \emph{LLM-based code edits}: the LLM generates targeted code edits constrained to the localized fault regions, producing minimal modifications that directly address the CI failure. The LLM is guided by a structured prompt to ensure minimal, localized, and semantically correct edits without introducing unrelated changes.

\begin{promptbox}[title={Prompt Template: Direct LLM-Based Patch Generation}]
\begin{verbatim}
Task:
Generate minimal code edits to repair localized CI faults.

Inputs:
- Fault localization data (code snippets and line ranges)
- File structure and CI failure context

Instructions:
1. Modify only the identified faulty regions.
2. Preserve all unrelated code and formatting.
3. Ensure syntactic and semantic correctness.
4. Return complete corrected code blocks for replacement.
5. If no modification is required, return an empty list.

Output:
[
  { "original_snippet": "...", "fixed_snippet": "..." }
]
\end{verbatim}
\end{promptbox}

(3) \emph{Patch construction and validation}: all changes are consolidated into a single patch and applied using a three-way merge to ensure conflict-free integration. Repair correctness is evaluated by re-executing the original CI configuration using the \benchmark evaluation setup. A repair is considered successful only if the CI pipeline passes without introducing new failures.

\section{Experiment Results}
\label{sec:result}
\subsection{RQ1: How effectively do LLMs repair CI failures?}

\phead{Motivation.}
Continuous Integration (CI) repair is a challenging software engineering task that requires interpreting noisy execution logs, identifying failure causes, localizing faults, and generating correct code modifications. Despite its practical importance, there is currently no standardized benchmark or baseline system for evaluating automated CI repair. While large language models (LLMs) have demonstrated strong performance on existing code-related benchmarks, their effectiveness on CI-style failures remains unclear.

To address this gap, we use \benchmark, a benchmark of real-world CI failures, together with a \textbf{reference agentic CI repair framework}. This setup enables controlled evaluation of whether CI artifacts contain sufficient information for automated systems to diagnose failures, localize repair-relevant code regions, and generate patches that restore CI success. It also enables a fair comparison of different LLM backbones under identical repair settings.

\phead{Approach.}
We evaluate the \textbf{reference CI repair framework} on all 567 CI failure instances in \benchmark (Section~\ref{sec:experiment}). Across all experiments, the repair pipeline is held constant, including CI artifact processing, log analysis, fault localization, patch generation workflow, prompts, and configuration settings. Only the underlying LLM backbone is varied, enabling a controlled comparison under identical repair conditions.

We assess both intermediate localization quality and end-to-end repair effectiveness. Let $\mathcal{D}$ denote the benchmark dataset and $N = |\mathcal{D}| = 567$. For each issue $i \in \mathcal{D}$, let $G_i$ denote the set of ground-truth modified files, and $R_i=\langle r_1,r_2,\dots\rangle$ the ranked localization outputs.

\smallskip
\uheadu{Top-$k$ Localization.}
Following prior fault localization studies, we evaluate whether repair-relevant files are ranked near the top of the predicted list. We report \textbf{Top-$k$} (for $k \in \{1, 3, 5\}$), defined as:
\[
\text{Top-}k=
\frac{1}{N}\sum_{i=1}^{N}
\mathbf{1}\!\left(G_i \cap \{r_1,\dots,r_k\}\neq\emptyset\right)\times100
\]
where $\mathbf{1}(\cdot)$ is the indicator function. Top-1 captures first-choice localization accuracy, while Top-3 and Top-5 reflect broader candidate coverage. We additionally report \textbf{Mean Average Precision (MAP)}, which summarizes overall ranking quality across benchmark issues.

\smallskip
\uheadu{Patch Application Coverage.}
We additionally report \textbf{Applied Patches}, defined as the number of benchmark issues for which the framework produced a patch that was successfully applied to the target repository and submitted to CI evaluation, regardless of final repair success. This metric captures how often the framework generates executable repair attempts before repository-level validation.

\smallskip
\uheadu{End-to-end Repair Success.}
We use \textbf{Pass@1} as the primary metric, defined as the percentage of issues successfully repaired on the first attempt after full repository-level CI validation. This metric captures whether the framework can jointly identify relevant repair targets, generate valid code edits, and produce repository-compatible fixes that satisfy the CI pipeline.

\phead{Results.}
\textbf{\textit{Across all four evaluated LLMs, Pass@1 ranges from 7.9\% to 18.9\%, with even the strongest model repairing fewer than one in five CI failures.}} Table~\ref{tab:rq1_main} reports end-to-end repair performance. GPT-5-mini achieves the highest repair success at 18.9\%, followed by DeepSeek-Coder (15.9\%), DeepSeek-Chat (13.2\%), and GPT-4o-mini (7.9\%). These results highlight the difficulty of repository-level CI repair for current LLMs. This difficulty reflects the nature of \benchmark, where repair correctness is determined by full CI re-execution rather than test-only oracles, and a candidate patch must satisfy every active validation stage in the original workflow.

\textbf{\textit{All four models achieve comparable fault localization, with Top-1 and MAP each varying by less than 4 points, yet their Pass@1 differs by more than 2 times.}} Table~\ref{tab:rq1_ranking} shows the localization results. GPT-4o-mini illustrates the gap most clearly: it matches DeepSeek-Coder on Top-1 (42.50\%) and slightly outperforms it on Top-3 (50.97\% vs. 50.44\%) and Top-5 (54.50\% vs. 52.38\%), yet its Pass@1 is roughly half that of DeepSeek-Coder (7.9\% vs. 15.9\%). This discrepancy indicates that identifying relevant files is necessary but not sufficient for successful CI repair. CI failures often involve multiple interacting components rather than a single file, so partially correct localization may omit additional required changes, while over-localization may introduce unnecessary modifications that break other parts of the pipeline. Successful repair thus requires not only identifying relevant files, but also determining the complete and minimal set of repository-consistent changes.

\textbf{\textit{Across all models, 73\% to 89\% of applied patches fail CI validation, revealing a large and consistent gap between successful patch application and full CI success.}} All four models produce patches that are successfully applied for the majority of benchmark issues (416 to 455 out of 567), yet only a much smaller subset subsequently passes full CI validation. For GPT-5-mini, 455 applied patches are evaluated but only 107 pass CI, meaning that roughly 76\% fail during validation. This pattern holds consistently across the other models. Because validation re-executes the complete CI pipeline rather than isolated tests, a patch must satisfy every active stage, including environment setup, dependency resolution, static analysis, and test execution, to be counted as a successful repair. A failure in any stage is sufficient to invalidate the patch, which explains the large drop between patch application and final CI success and reinforces that \benchmark evaluates repair correctness under stricter conditions than test-centric benchmarks.

\begin{table}[t]
\centering
\footnotesize
\caption{Overall CI repair performance on \benchmark. Pass@1 is the percentage of benchmark issues successfully repaired on the first attempt after full CI validation.}
\label{tab:rq1_main}
\begin{tabular}{lcc}
\toprule
\textbf{Model} & \textbf{Applied Patches} & \textbf{Pass@1} \\
\midrule
GPT-5-mini      & \textbf{455} & \textbf{18.9} \\
DeepSeek-Coder  & 438 & 15.9 \\
DeepSeek-Chat   & 416 & 13.2 \\
GPT-4o-mini     & 420 & 7.9 \\
\bottomrule
\end{tabular}
\end{table}

\begin{table}[t]
\centering
\footnotesize
\caption{Recall at Top-N localization performance on \benchmark. Recall@k measures whether at least one ground-truth repair-relevant file appears within the top-k ranked predictions. MAP summarizes the overall ranking quality across benchmark issues.}
\label{tab:rq1_ranking}
\begin{tabular}{lcccc}
\toprule
\textbf{Model} & \textbf{Top-1} & \textbf{Top-3} & \textbf{Top-5} & \textbf{MAP} \\
\midrule
GPT-5-mini      & \textbf{45.33} & \textbf{53.97} & \textbf{58.55} & \textbf{42.53} \\
DeepSeek-Coder  & 42.50 & 50.44 & 52.38 & 38.83 \\
GPT-4o-mini     & 42.50 & 50.97 & 54.50 & 38.26 \\
DeepSeek-Chat   & 41.62 & 47.27 & 48.15 & 37.92 \\
\bottomrule
\end{tabular}
\end{table}


\begin{tcolorbox}[colback=gray!10, colframe=black, boxrule=0.5pt, arc=2pt]
\textbf{Takeaway.}
Automated CI repair on \benchmark is challenging for current LLMs, with Pass@1 ranging from 7.9\% to 18.9\%. Although fault localization performance is similar across all models, Pass@1 differs by more than 2x, showing that identifying relevant files is necessary but not sufficient. Successful repair requires generating repository-consistent changes that satisfy all active CI validation stages.
\end{tcolorbox}




\subsection{RQ2: What is the impact of log analysis strategy on CI repair performance?}
\label{sub:overall_performance_log_analysis}

\phead{Motivation.} CI repair depends on extracting reliable failure evidence from CI logs, which are typically lengthy, noisy, and distributed across multiple execution stages. Relevant signals are often separated across distant log segments, and reported symptoms do not always reveal the underlying fault locations. Consequently, weak log understanding propagates to downstream stages and reduces repair effectiveness. In this RQ, we investigate how alternative log analysis strategies affect end-to-end CI repair quality and computational efficiency.

\phead{Approach.}
We compare two log analysis strategies within the same reference CI repair framework on all benchmark instances. For a controlled comparison, only the log analysis stage is changed, while all downstream components remain identical, including fault localization, patch generation, prompts, model backbone, and evaluation settings.

\uheadu{Agent-based Analysis(LLM).}
We use the default log analysis strategy described in Section~\ref{sec:experiment}. The framework applies an iterative LLM-based analysis process that incrementally inspects CI artifacts, reasons over intermediate findings, and constructs a structured failure summary for downstream localization and repair.

\uheadu{Retrieval-based Analysis (BM25).}
As an alternative, we evaluate a lightweight retrieval-based strategy using BM25. Retrieval-based context selection has been effective in prior software engineering tasks for efficiently identifying relevant artifacts from large search spaces~\citep{jimenez2024swebench,rafi2025graphretrievalfl}, motivating its evaluation for CI repair. CI logs are segmented into chunks, and BM25 retrieves failure-relevant log segments together with potentially related repository files using lexical overlap. The retrieved evidence is aggregated and provided to the LLM to generate a structured failure summary for downstream localization and patch generation.

\phead{Evaluation Metrics.}
We use the same repair metrics as in RQ1. To evaluate downstream localization quality, we report \textbf{Top-1}, \textbf{Top-3}, and \textbf{MAP}. Top-$k$ measures whether at least one ground-truth repair-relevant file appears within the top-$k$ ranked predictions, while MAP summarizes the overall ranking quality across benchmark issues. We report \textbf{Pass@1} as the primary end-to-end repair metric.

To quantify the effect of changing the log analysis strategy, let $P_{\text{agent}}$ and $P_{\text{retrieval}}$ denote the Pass@1 scores under the two settings. We compute the repair impact of Retrieval as:

\[
\text{Relative Drop (\%)}=
\frac{P_{\text{agent}}-P_{\text{retrieval}}}{P_{\text{agent}}}\times100
\]

Higher values indicate larger degradation relative to the default Agent-based strategy.

We further evaluate computational efficiency using execution traces collected for each benchmark issue. For each run, we record input tokens, output tokens, total tokens, API cost, and API calls, partitioned across log analysis, fault localization, and patch generation stages.

\phead{Results.} 
\textbf{\textit{Replacing Agent-based log analysis with BM25-based Retrieval consistently reduces repair success across all evaluated models, with relative Pass@1 drops ranging from 28.0\% to 75.9\%.}} Table~\ref{tab:rq2_main} reports the impact of log analysis strategy on end-to-end CI repair. When Agent-based analysis is replaced by BM25-based Retrieval, GPT-5-mini declines from 18.9\% to 10.6\%, DeepSeek-Coder from 15.9\% to 9.9\%, DeepSeek-Chat from 13.2\% to 9.5\%, and GPT-4o-mini from 7.9\% to 1.9\%. The consistent degradation across all four LLMs indicates that effective log analysis is a core requirement for successful CI repair, rather than a model-specific advantage.


\textbf{\textit{BM25-based Retrieval log analysis substantially reduces downstream localization quality across all evaluated models, with Top-1 declining by 24.0 to 37.7 points and MAP declining by 18.3 to 31.5 points.}} As shown in Table~\ref{tab:rq2_main}, this degradation appears for every model, indicating that the effect is systematic rather than model-specific. Even the strongest model, GPT-5-mini, experiences major performance losses, falling to roughly half of its Agent-based Top-1 and MAP levels under BM25-based Retrieval. GPT-4o-mini is affected most severely, with Top-1 dropping to 4.76\%, meaning that only a small fraction of issues place a repair-relevant file at the highest-ranked position. DeepSeek-Coder and DeepSeek-Chat follow the same overall trend despite differences in absolute repair strength. These results suggest that BM25-based Retrieval often captures surface lexical matches but fails to recover the richer failure evidence needed to accurately rank repair-relevant files. As a result, weaker upstream log understanding propagates to later stages, leading to poorer localization and lower end-to-end repair success.

\begin{table}[t]
\centering
\footnotesize
\caption{Impact of log analysis strategy on ranking-based fault localization and CI repair performance. Recall@k measures whether at least one ground-truth repair-relevant file appears within the top-k ranked predictions. MAP summarizes the overall ranking quality across benchmark issues. Relative Pass@1 drops of the Retrieval setting compared with the default Agent-based strategy are shown in parentheses.}
\label{tab:rq2_main}
\begin{tabular}{l l c c c c}
\toprule
\textbf{Model} & \textbf{Strategy} & \textbf{Top-1} & \textbf{Top-3} & \textbf{MAP} & \textbf{Pass@1} \\
\midrule
GPT-5-mini     
& Agent     & \textbf{45.33} & \textbf{53.97} & \textbf{42.53} & \textbf{18.9} \\
& Retrieval & 19.40 & 27.69 & 21.52 & 10.6 (-43.9\%) \\
\midrule

DeepSeek-Coder 
& Agent     & 42.50 & 50.44 & 38.83 & 15.9 \\
& Retrieval & 17.28 & 23.28 & 18.67 & 9.9 (-37.7\%) \\
\midrule

DeepSeek-Chat  
& Agent     & 41.62 & 47.27 & 37.92 & 13.2 \\
& Retrieval & 17.64 & 24.16 & 19.66 & 9.5 (-28.0\%) \\
\midrule

GPT-4o-mini    
& Agent     & 42.50 & 50.97 & 38.26 & 7.9 \\
& Retrieval & 4.76 & 7.41 & 6.72 & 1.9 (-75.9\%) \\
\bottomrule
\end{tabular}
\end{table}

\begin{table*}[t]
\centering
\footnotesize
\caption{Stage-wise efficiency comparison of log analysis strategies using GPT-5-mini. Totals are aggregated over all benchmark issues for each strategy. Token counts are reported in millions (M). Cost is the total API expenditure in USD (\$), and Calls denotes the number of LLM API requests made per stage.}
\label{tab:rq2_efficiency}
\resizebox{\textwidth}{!}{%
\begin{tabular}{l ccccc ccccc}
\toprule
& \multicolumn{5}{c}{\textbf{Agent (LLM)}} & \multicolumn{5}{c}{\textbf{Retrieval (BM25)}} \\
\cmidrule(lr){2-6} \cmidrule(lr){7-11}
\textbf{Stage}
& \textbf{Input} & \textbf{Output} & \textbf{Total} & \textbf{Cost (\$)} & \textbf{Calls}
& \textbf{Input} & \textbf{Output} & \textbf{Total} & \textbf{Cost (\$)} & \textbf{Calls} \\
\midrule
Log Analysis
& 99.47 & 11.47 & 110.94 & 159.88 & 3,704
& 11.53 & 1.88 & 13.41 & 20.96 & 408 \\
Fault Localization
& 54.23 & 8.45 & 62.68 & 96.83 & 9,042
& 12.27 & 3.61 & 15.88 & 29.36 & 3,861 \\
Patch Generation
& 17.53 & 4.35 & 21.88 & 38.41 & 3,362
& 3.81 & 1.71 & 5.52 & 11.71 & 984 \\
\midrule
Total
& 171.23 & 24.27 & 195.49 & 295.12 & 16,108
& 27.61 & 7.20 & 34.81 & 62.03 & 5,253 \\
\bottomrule
\end{tabular}
}
\end{table*}

\textbf{\textit{Across the full CI repair pipeline, Agent-based analysis consumes substantially more computational resources than BM25-based Retrieval, requiring 5.6$\times$ more total tokens, 4.8$\times$ higher API cost, and 3.1$\times$ more API calls.}} Table~\ref{tab:rq2_efficiency} summarizes the stage-wise efficiency comparison. Agent-based analysis consumes 195.49M tokens at \$295.12 and 16,108 API calls overall, while BM25-based Retrieval completes the same workload at roughly one-fifth of that budget (34.81M tokens, \$62.03, 5,253 API calls). The gap is most pronounced in log analysis, where Agent-based analysis alone spends 110.94M tokens, more than 8$\times$ BM25-based Retrieval's 13.41M. This reflects how the two strategies operate: Agent-based analysis iteratively inspects logs, refines hypotheses, and consolidates evidence into a structured failure explanation, whereas BM25-based Retrieval relies on lexical matching through keyword overlap. As a result, BM25-based Retrieval is efficient but may miss semantically relevant evidence or surface misleading artifacts, and this weaker diagnosis carries into later stages. Fault localization then drops to a quarter of Agent-based cost (15.88M vs. 62.68M tokens), and patch generation follows the same trend (5.52M vs. 21.88M), since both stages operate on lower-quality context. Overall, iterative reasoning yields stronger repair outcomes, despite its higher token usage, API cost, and API calls.

\begin{tcolorbox}[colback=gray!10, colframe=black, boxrule=0.5pt, arc=2pt]
\textbf{Takeaway.}
Log analysis strategy is a major factor in CI repair success on \benchmark. Replacing Agent-based reasoning with BM25-based Retrieval reduces Pass@1 by 28.0\% to 75.9\% across all evaluated LLMs, while also lowering Top-1 localization by 24.0 to 37.7 points. Although Agent-based analysis requires higher resources (5.6$\times$ more tokens and 4.8$\times$ higher cost), the consistent gains in localization and repair effectiveness show that accurate failure diagnosis is more valuable than lightweight retrieval efficiency for artifact-rich CI repair tasks.
\end{tcolorbox}

\subsection{RQ3: How does CI repair performance vary across failure types?}
\label{sub:overall_performance_ci_error_type}

\phead{Motivation.}
CI failures originate from diverse sources that differ in locality, observability, and the reasoning required to resolve them. A formatting 
violation, a missing dependency, and a misconfigured environment all surface as a failed CI run, but each demands a fundamentally different repair strategy. Aggregate repair rates, therefore, obscure where current systems actually succeed and where they fail. Disaggregating performance by failure type is essential for identifying which categories LLM-based repair can reliably handle today, and which expose capability gaps that future work must address.

\phead{Approach.}
We categorize CI failures into 12 types following the diagnostic signals listed in Table~\ref{tab:failure_type_definitions}, spanning code-level issues (formatting, linting, syntax, runtime, testing, assertions, type checking), project-level issues (dependencies, package installation, configuration), and environment-level issues, as well as documentation. Because a single CI workflow may execute multiple validation steps, a single instance can carry multiple failure-type labels. We assign labels by combining validation failures observed in CI logs with the ground-truth patches that resolved them. Since a CI instance is considered repaired only when all validation steps pass, a successful repair implies that every co-occurring failure type was resolved. This labeling scheme lets us evaluate per-category performance while preserving the multi-failure structure of real CI workflows.

\begin{table}[H]
\caption{Definitions of the 12 CI failure types in 
\benchmark, grouped by origin. Diagnostic signals indicate the typical 
log evidence used during labeling.}
\label{tab:failure_type_definitions}
\centering
\footnotesize
\scalebox{0.9}{
\setlength{\tabcolsep}{0.3cm}
\renewcommand{\arraystretch}{1.15}
\begin{tabular}{p{2.6cm} p{4.2cm} p{5.2cm}}
\toprule
\textbf{Failure Type} & \textbf{Definition} & \textbf{Typical Diagnostic Signal} \\
\midrule
\multicolumn{3}{l}{\textit{Code-level issues}} \\
\midrule
Code Formatting & Violations of automated formatting rules (e.g., spacing, indentation, import order). & Output from \texttt{black}, \texttt{ruff format}, \texttt{isort}; ``would reformat'' or formatter exit code. \\
Code Linting & Violations of static analysis rules for code quality or style. & Output from \texttt{ruff}, \texttt{flake8}, \texttt{pylint} with rule codes (e.g., \texttt{E501}, \texttt{F401}). \\
Syntax Error & Code that fails to parse. & \texttt{SyntaxError} or parser failure during import or compilation. \\
Runtime Error & Uncaught exceptions raised during execution outside of test assertions. & Tracebacks ending in exceptions such as \texttt{TypeError}, \texttt{AttributeError}, \texttt{KeyError}. \\
Test Failure & A test executes but does not produce the expected outcome. & \texttt{pytest} or \texttt{unittest} report with failed test cases. \\
Assertion Error & A test fails specifically due to an \texttt{assert} statement. & \texttt{AssertionError} in the test traceback. \\
Type Checking & Violations reported by a static type checker. & Errors from \texttt{mypy}, \texttt{pyright}, or equivalent. \\
\midrule
\multicolumn{3}{l}{\textit{Project-level issues}} \\
\midrule
Dependency Issues & Conflicts or inconsistencies between declared and resolved dependencies (e.g., version incompatibilities, unused or deprecated packages in \texttt{requirements.txt}). & Resolver errors, version conflicts, incompatible constraints in \texttt{pip}/\texttt{poetry} output; \texttt{ImportError} or \texttt{ModuleNotFoundError} at runtime. \\

Pkg.\ Install Error & Failures during the physical installation of a package after dependency resolution. & \texttt{pip install} errors, build failures for native extensions, missing or incompatible wheels, missing system-level build dependencies. \\

Configuration Error & Invalid or inconsistent project settings unrelated to dependency declarations, including misconfigured Python version targets. & Errors parsing \texttt{pyproject.toml} or \texttt{setup.cfg}; misreferenced paths or options; \texttt{python\_requires} constraints conflicting with installed packages. \\
\midrule
\multicolumn{3}{l}{\textit{Environment-level issues}} \\
\midrule
Environment Error & Failures originating from the runner or execution context. & Missing system libraries, unsupported runner or interpreter versions, unavailable external services. \\
\midrule
\multicolumn{3}{l}{\textit{Documentation}} \\
\midrule
Doc/Docstring & Failures in documentation-build or docstring formatting and validation steps. Includes violations flagged by docstring formatters. & Errors from \texttt{sphinx-build}, docstring linters, \texttt{docformatter} (reformatting violations), or doc-test runners. \\
\bottomrule
\end{tabular}
}
\end{table}

\phead{Results.} \textbf{\textit{Repair success follows a consistent three-tier 
pattern across all models and both log analysis strategies.}} 
Table~\ref{tab:ci_failure_by_model_two_strategies} reveals a stable hierarchy of repairability that holds regardless of model capability. The top tier comprises code formatting and linting, which reach success rates of roughly 8 to 35.5 percent. Under the strongest configuration (GPT-5-mini 
with Agent), Code Formatting reaches 35.5 percent (43/121) and Code Linting 
reaches 17.8 percent (37/208). The middle tier covers test failures, runtime errors, and syntax errors, where success rates are moderate but uneven; GPT-5-mini repairs 13.0 percent of test failures (15/115), 10.6 percent of runtime errors (9/85), and 10.3 percent of syntax errors (7/68). The bottom tier consists of package installation, configuration, and environment failures, which remain at 0 to 8.8 percent in nearly every setting, with many categories producing zero successful repairs even under the strongest model. This stratification is robust: it appears across all four LLMs and persists under both Agent and Retrieval log analysis.

\begin{table}[H]
\caption{
Results of the reference CI repair framework on \benchmark.
\#Occur.\ denotes the number of failure-type occurrences (one instance may contribute multiple labels).
\textbf{Agent} and \textbf{Retrieval} denote two CI log analysis strategies.
}
\label{tab:ci_failure_by_model_two_strategies}
\centering
\footnotesize
\setlength{\tabcolsep}{3.5pt}
\renewcommand{\arraystretch}{1.1}
\resizebox{\textwidth}{!}{%
\begin{tabular}{l c cc cc cc cc}
\toprule
\multirow{2}{*}{\textbf{CI Failure Type}} &
\multirow{2}{*}{\textbf{\#Occur.}} &
\multicolumn{2}{c}{\textbf{GPT-5-mini}} &
\multicolumn{2}{c}{\textbf{DeepSeek-Coder}} &
\multicolumn{2}{c}{\textbf{DeepSeek-Chat}} &
\multicolumn{2}{c}{\textbf{GPT-4o-mini}} \\
\cmidrule(lr){3-4}\cmidrule(lr){5-6}\cmidrule(lr){7-8}\cmidrule(lr){9-10}
& & \textbf{Agent} & \textbf{Retrieval}
  & \textbf{Agent} & \textbf{Retrieval}
  & \textbf{Agent} & \textbf{Retrieval}
  & \textbf{Agent} & \textbf{Retrieval} \\
\midrule
\multicolumn{10}{c}{\textbf{(a) Failure-type-level CI Repair}} \\
\midrule
Code Linting        & 208 & 37 & 24 & 36 & 23 & 29 & 22 & 17 & 2 \\
Dependency Issues   & 148 & 10 & 4  & 6  & 3  & 2  & 1  & 6  & 0 \\
Code Formatting     & 121 & 43 & 21 & 33 & 19 & 25 & 22 & 17 & 9 \\
Test Failure        & 115 & 15 & 11 & 11 & 9  & 12 & 6  & 4  & 1 \\
Runtime Error       & 85  & 9  & 7  & 14 & 7  & 12 & 6  & 2  & 0 \\
Syntax Error        & 68  & 7  & 1  & 10 & 1  & 5  & 1  & 7  & 0 \\
Pkg. Install Error  & 48  & 0  & 0  & 1  & 0  & 1  & 0  & 0  & 0 \\
Configuration Error & 34  & 1  & 0  & 3  & 0  & 2  & 1  & 2  & 0 \\
Environment Error   & 32  & 1  & 0  & 1  & 0  & 3  & 1  & 1  & 0 \\
Assertion Error     & 28  & 2  & 2  & 1  & 2  & 1  & 1  & 0  & 0 \\
Doc/Docstring       & 4   & 0  & 0  & 0  & 0  & 0  & 0  & 0  & 0 \\
Type Checking       & 2   & 1  & 0  & 1  & 0  & 1  & 1  & 1  & 0 \\
\textbf{Total}      & \textbf{893} & 126 & 70 & 117 & 64 & 93 & 62 & 57 & 11 \\
\bottomrule
\end{tabular}%
}
\end{table}

\textbf{\textit{Stronger models and better log analysis amplify performance on already tractable categories rather than unlocking the hardest ones.}} Moving from GPT-4o-mini to GPT-5-mini under Agent analysis roughly doubles total repairs (57 to 126), but the gains concentrate in formatting and linting. The hardest categories barely move. Package Installation Failures top out at 2.1 percent (1/48) across all settings, Configuration Errors peak at 8.8 percent (3/34), and Environment Errors peak at 9.4 percent (3/32). Even when overall repair counts rise sharply with stronger backbones, the bottom tier remains stuck. This indicates that current advances primarily strengthen performance on failures the system can already partially handle, while structurally complex CI failures stay largely unresolved.

\textbf{\textit{The performance gap reflects differences in 
failure structure, not just difficulty.}} The categories that succeed share a common property: error messages explicitly identify both the violation and its location, enabling a near-direct mapping from log signal to code edit. The categories that fail share the opposite property: their signals 
are indirect, distributed, or require external context. A version conflict in a dependency log does not reveal which constraint to relax; a missing package error does not specify which configuration file to update; an environment failure may require coordinated changes across runner configuration, dependency specifications, and source code. Resolving these 
failures demands repository-level reasoning over project structure, toolchains, and inter-component consistency, which single-pass repair cannot reliably provide.

\begin{tcolorbox}[colback=gray!10, colframe=black, boxrule=0.5pt, arc=2pt]
\textbf{Takeaway.}
Repair success is driven by signal locality, not model strength. Formatting and linting, which carry explicit error locations, reach 35.5\% and 17.8\% under the best configuration. Dependency, configuration, and environment 
failures stay below 9\% across all 8 model-strategy combinations, with package installation peaking at 2.1\%. Closing this gap requires repository-aware reasoning, dependency analysis, and iterative validation, 
not just stronger backbones.
\end{tcolorbox}

\section{Discussion}
\label{sec:discussion}
\phead{Error Taxonomy.}
To understand the failure characteristics exposed by \benchmark, we analyze unsuccessful repairs based on the underlying limitations of \textsc{LLM-based CI repair} rather than surface-level CI error labels. A major source of failure is \emph{log-bounded reasoning}. CI logs usually expose only the first failing step and provide limited context, which leads models to focus on the reported error location even when the root cause lies in dependent files, other pipeline stages, or hidden configuration assumptions. Another common failure mode involves \emph{cross-file and dependency-driven errors}. Many test and runtime failures require reasoning across multiple files, caller-callee relationships, or shared dependencies, which are rarely explicit in CI logs and are difficult for LLM-based solutions to infer without program analysis.

Configuration-related failures further reduce repair success, as they require project-specific knowledge such as environment variables, dependency constraints, and repository layout, often combined with iterative validation. Such information is typically missing from log-based analysis and static code snapshots. Finally, CI pipelines follow a fail-fast execution model. Even when a reported failure is fixed, additional errors may remain hidden and only surface after re-execution. Overall, these failure modes indicate that effective CI repair requires reasoning beyond logs, combining code structure, dependencies, configuration, and execution feedback, which \benchmark is designed to reveal through error-based analysis.

\phead{Limitations and Challenges.}
\benchmark has several limitations and also highlights broader challenges for automated CI repair. First, only a limited number of CI instances could be reliably collected. Many workflows are not reproducible due to evolving dependencies, external tooling, changing runner environments, missing or unavailable logs, maintenance or infrastructure-focused pipelines, and repeated identical failures across runs. Together, these factors substantially reduce the number of analyzable and reproducible workflows.

Second, CI logs provide only partial observability, as workflows often terminate at the first failing step, preventing later stages from producing diagnostic information until earlier issues are resolved. In addition, CI behavior is not always fully reproducible across forks or over time, leading to non-deterministic outcomes even for identical configurations. Identifying reliable ground-truth repairs from version control history is also challenging, as multiple intermediate commits may exist between failing and passing states and often include unrelated changes or refactorings, making it difficult to isolate the precise fix. Finally, while we evaluate a representative CI repair pipeline, many CI failures require deeper environment modeling, dependency resolution, and iterative execution reasoning beyond the capabilities of single-pass LLMs, highlighting the need for future agent-based CI repair systems with richer tooling.

\section{Threats to Validity}
\label{sec:threats}
\phead{Internal Validity.}
Our evaluation involves large language models whose outputs are inherently non-deterministic. Even under identical inputs and configurations, models may produce variations in content, structure, and formatting. Such variability can affect intermediate stages of the CI repair pipeline, including failure interpretation, candidate file selection, and patch generation. As these stages are interdependent, small differences in early outputs may propagate and lead to different repair outcomes. In addition, CI workflows may exhibit non-deterministic behavior due to environment variability and external dependencies, which can further affect evaluation results.

\phead{Construct Validity.}
We evaluate repair effectiveness using Pass@1, defined as the proportion of generated patches that result in successful CI execution. While this metric reflects end-to-end repair success, it does not consider partial fixes that resolve only a subset of validation failures. Furthermore, the evaluation of generated patches is inherently guided by observable failure signals in CI logs. As these logs typically expose only the first failing validation step, they may not fully represent the underlying root causes, which can affect both patch generation and the interpretation of repair success.

\phead{External Validity.}
Our benchmark is constructed from Python-based repositories; however, the underlying methodology is not inherently language-specific. It relies on CI failure–success pairs, execution logs, workflow configurations, and ground-truth patches, which are common across CI systems. In principle, the same approach can be extended to other programming languages with comparable CI workflows. However, differences in ecosystem-specific tooling, dependency management, and CI configurations may affect reproducibility and failure characteristics, requiring further validation to assess generalizability.

\phead{Reproducibility and Dataset Limitations.}
Reconstructing CI failures is inherently challenging due to the evolving nature of execution environments, dependencies, and tooling. Changes such as version updates, deprecations, or modifications in external services can prevent previously observed CI failures from being reproduced reliably over time. Consequently, maintaining reproducible CI-based benchmarks requires continuous maintenance of workflows and execution environments to preserve compatibility with the original failure conditions. In addition, identifying ground-truth repairs from version histories may introduce noise, as commits can include unrelated modifications alongside the actual fix. To ensure reproducibility, we construct ground-truth patches by retaining only changes relevant to the observed CI failure. In some cases, additional adjustments are required due to differences in the reconstructed CI environment (e.g., updated dependencies or configuration changes). Consequently, the resulting ground-truth patches may not exactly match the full set of changes between the original failing and succeeding commits, but instead represent an approximation of the minimal modifications required to reproduce a successful CI execution.

\section{Conclusion}
\label{sec:conclusion}
We presented \benchmark, a repository-aware benchmark for evaluating automated program repair under realistic CI workflows. It validates repairs by re-executing complete CI pipelines on GitHub Actions, requiring candidate patches to succeed across heterogeneous CI stages beyond test execution. By mining real failure-success CI episodes from actively maintained repositories and organizing them into multi-label CI error categories, the benchmark captures the complexity of real-world CI failures. Our evaluation of LLM-based repair systems shows strong performance on localized, tool-enforced failures but limited success on environment- and dependency-driven failures, highlighting both the promise and current limitations of automated CI repair. We hope \benchmark serves as a stable and extensible platform for future CI-native repair systems that reason jointly over code, workflows, environments, and execution feedback.


\section{Declarations}

\subsection*{Funding}
This research received no external funding.

\subsection*{Ethical Approval}
Not applicable.

\subsection*{Informed Consent}
Not applicable.

\subsection*{Author Contributions}
The contributions of the authors are as follows.

Rabeya Khatun Muna: Conceptualization, Methodology, Software, Data curation, Formal analysis, Investigation, Benchmark construction, Experiments, Writing - original draft.

Md Nakhla Rafi: Software, Formal analysis, Validation, Investigation, Writing - review and editing.

Tse-Hsun (Peter) Chen: Conceptualization, Supervision, Funding acquisition, Writing - review and editing.

\subsection*{Data Availability}
All artifacts required to reproduce the experiments are publicly available. The CI-Repair-Bench benchmark, including baseline implementations, evaluation scripts, and reproduction instructions, is available at: \url{https://github.com/RabeyaMuna/CI-REPAIR-BENCH}. The dataset containing benchmark instances, metadata, CI logs, patches, and annotations is publicly available at: \url{https://huggingface.co/datasets/ci-benchmark-user/ci-repair-bench}.

\subsection*{Conflict of Interest}
The authors declare that they have no conflict of interest.

\subsection*{Code Availability}
The source code used in this study is publicly available through the project replication package and GitHub repository listed above.


\appendix
\section{Dataset Statistics and Distribution}
\label{appendix:dataset_stats}

This appendix provides detailed per-repository statistics for \benchmark, including the number of CI failure instances and associated error categories for each repository. These distributions complement the aggregate dataset summary presented in Section~\ref{ref:dataset_summary}. Table~\ref{tab:repo_stats} reports per-repository instance counts and observed error types.

\begin{table}[t]
\centering
\caption{Per-repository dataset statistics (CI failure instances and associated error types).}
\label{tab:repo_stats}
\footnotesize
\setlength{\tabcolsep}{4pt}
\renewcommand{\arraystretch}{1.1}
\resizebox{\textwidth}{!}{%
\begin{tabular}{l r l}
\toprule
\textbf{Repository} & \textbf{Inst.} & \textbf{Error Types} \\
\midrule
agno & 84 & Format, Lint, Deps, Runtime, Syntax, Test, Env \\
axolotl & 51 & Lint, Deps, Runtime, Test \\
litellm & 30 & Lint, Deps, Runtime, Syntax, Test \\
conan & 28 & Deps, Runtime, Syntax, Test, Env \\
flower & 27 & Format, Lint, Runtime \\
browser-use & 27 & Lint, Deps, Runtime, Test \\
taipy & 19 & Lint, Deps, Runtime, Test \\
docsgpt & 15 & Format, Runtime, Deps \\
camel & 15 & Format, Runtime \\
aiohttp & 14 & Lint, Deps, Runtime, Syntax, Test \\
crewai & 14 & Runtime, Syntax, Test \\
maxkb & 12 & Format, Runtime \\
aider & 9 & Runtime, Deps, Test \\
django-import-export & 8 & Format, Deps, Runtime \\
agentscope & 7 & Lint, Deps, Runtime \\
aisuite & 7 & Assert, Runtime, Deps \\
beets & 6 & Format, Lint \\
cloud-init & 6 & Deps, Runtime, Test \\
calibre & 6 & Format, Deps, Runtime \\
aws-cli & 5 & Deps, Runtime \\
dask & 5 & Runtime, Deps, Test \\
httpx & 5 & Format, Lint, Runtime \\
skypilot & 5 & Format \\
sqlglot & 5 & Format, Lint, Syntax, Test \\
airbyte & 4 & Format, Lint \\
starlette & 4 & Syntax, Test \\
uvicorn & 4 & Syntax, Runtime, Test \\
deepcode & 4 & Format, Lint \\
accelerate & 4 & Format \\
diffusers & 4 & Format, Lint, Deps \\
kitty & 4 & Format, Runtime, Deps \\
openai-python & 4 & Assert, Deps, Runtime \\
cpython & 4 & Format, Lint, Runtime \\
LightRAG & 3 & Format \\
olmocr & 3 & Format, Runtime \\
lm-evaluation-harness & 3 & Format, Lint \\
facefusion & 3 & Lint, Runtime, Test \\
integration & 3 & Format \\
lightrag & 3 & Format, Lint \\
locust & 3 & Format, Runtime, Test \\
\bottomrule
\end{tabular}%
}
\end{table}

\section{LLM Models and Versions}
\label{appendix:model_versions}

To ensure reproducibility, we explicitly document the language models used in our evaluation. OpenAI models are referenced by their official API identifiers and fixed snapshots when available, while DeepSeek models are referenced by the identifiers exposed through the DeepSeek API, as detailed snapshot-level versioning is not fully disclosed by the vendor.

\begin{itemize}
\item \textbf{GPT-5 mini}: OpenAI API model \texttt{gpt-5-mini} (fixed snapshot: \texttt{gpt-5-mini-2025-08-07}), a faster, cost-efficient variant of the GPT-5 family released alongside GPT-5 on August 7, 2025~\citep{openai_gpt5_2025, openai_gpt5mini_2025}. It provides a 128k-token context window optimized for structured reasoning and code-related tasks.
  
  \item \textbf{GPT-4o-mini}: OpenAI API model \texttt{gpt-4o-mini}, a compact multimodal model optimized for high-throughput and low-latency text generation~\citep{openai_gpt4o_mini}. We use the fixed snapshot \texttt{gpt-4o-mini-2024-07-18}, which provides a 128k-token context window and supports structured outputs.
  
  \item \textbf{DeepSeek-Coder}: DeepSeek code-oriented model accessed via the DeepSeek API, designed for program synthesis and code understanding tasks~\citep{deepseek_coder_2024}. The model identifier follows the vendor-provided naming convention; explicit snapshot-level versioning is not publicly documented.
  
  \item \textbf{DeepSeek-Chat (v3)}: DeepSeek conversational model (v3) accessed via the DeepSeek API, intended for general-purpose dialogue and reasoning~\citep{deepseek_v3_2024}. As with DeepSeek-Coder, detailed snapshot-level versioning information is not publicly available.
\end{itemize}

Unless otherwise stated, all models were invoked using the latest stable API releases available at the time of experimentation, with default decoding parameters and without task-specific fine-tuning.

\bibliographystyle{spbasic}
\bibliography{references}
\end{document}